\documentclass{imag-ms-template}

\title{Benchmarking External Generalization of SPD Matrix Learning for Resting-State fMRI Connectome Prediction}

\author{Ce Ju$\dag$, Antoine Collas$\dag$, Florent Bouchard$\ddag$, Bertrand Thirion$\dag$\\
{\small $\dag$ Inria, CEA, Université Paris-Saclay, Palaiseau, France}\\
{\small $\ddag$ Université Paris-Saclay, CNRS, CentraleSupélec, L2S, France}\\
{\small Correspondence:\{ce.ju,bertrand.thirion\}@inria.fr}
}

\date{August 31, 2026}

\usepackage{booktabs}
\usepackage{microtype}
\usepackage{url}
\usepackage{tabularx}
\usepackage{float}
\hypersetup{hidelinks}
\begin{document} 
\raggedbottom

\maketitle


\begin{abstract}
Resting-state functional magnetic resonance imaging (rs-fMRI) functional connectivity (FC) matrices are widely used for individual-level prediction, but strong performance within one cohort may not generalize to a new cohort. We ask whether within-dataset performance remains when the test data come from an entirely held-out rs-fMRI dataset. Each scan is represented as a regularized symmetric positive definite (SPD) correlation connectome, which allows methods to use the geometry of the SPD manifold. We introduce a reproducible age-prediction benchmark across six rs-fMRI datasets: COBRE, ADNIDOD, Cam-CAN, ABIDE, OASIS-3, and ADNI. The benchmark compares a vectorized correlation baseline, Tangent-Space Ridge, SPDNet, and split-wise Riemannian harmonization under within-dataset GroupKFold, pooled GroupKFold, and leave-one-dataset-out (LODO) evaluation. Within-dataset and pooled GroupKFold results are substantially more favorable than LODO results. When an entire dataset is held out, prediction error increases, differences among methods narrow, and performance is strongly affected by age-range mismatch and cohort heterogeneity. The benchmark provides common inputs, model settings, data splits, and analysis scripts so that future SPD matrix learning methods can be evaluated under the same external-validation protocol.
\end{abstract}

\section{Introduction}

Resting-state fMRI (rs-fMRI) is widely used to assess large-scale brain organization. Functional connectivity (FC) matrices built from regional time series are also widely used for individual prediction tasks such as brain age estimation~\cite{smith2013functional,dadi2019benchmarking,pervaiz2020optimising}. A model may appear convincing when the training and test data come from the same cohort. However, that does not guarantee that it will perform well with the next cohort. A more difficult question is whether that performance will hold when the cohort changes. This question is important because neuroimaging prediction is most useful when it generalizes beyond the dataset on which it was developed. Scanner and site effects, denoising choices, age range, diagnosis mix, and implementation details all influence the prediction problem. Therefore, cross-validation within a dataset can give an answer that appears stronger than it really is. While it can demonstrate that a model functions within a specific evaluation setting, it cannot independently demonstrate whether the model has captured a generalizable connectome signal or if performance has been influenced by the evaluation setting~\cite{varoquaux2017assessing,varoquaux2018cross}.

External cohort validation is therefore not a secondary check for large-scale fMRI prediction; the same principle applies broadly to medical data analysis~\cite{varoquaux2023evaluating}. It is the test that separates a model that fits one collection from a pipeline that can be carried to the next study. A benchmark with fixed splits, shared inputs, and simple reference models can therefore serve as a reference point for future large-scale rs-fMRI analyses, including models trained on many cohorts~\cite{poldrack2019computational}.

Symmetric positive definite (SPD) matrix learning provides a natural framework for covariance and correlation matrices, and Riemannian methods are now widely used in neuroimaging and related domains~\cite{varoquaux2010detection,huang2017riemannian,ju2025spd}. For this reason, SPD matrix learning is well suited to this benchmark. The input is a regularized FC matrix, and SPD methods are designed to use this matrix structure directly. As Riemannian geometry has become an important way to model covariance and correlation matrices in neuroimaging, the benchmark investigates whether these methods still generalize when the test cohort is new. If gains are measured mainly within one dataset, or under pooled cross-validation where related cohorts contribute to both training and testing, the results remain useful but incomplete as evidence of robustness across cohorts. The field therefore needs external validation benchmarks that use shared connectome construction, shared splits, and comparable model implementations.

We address this need with a benchmark across six datasets for rs-fMRI connectome regression. The benchmark includes COBRE, ADNIDOD, Cam-CAN, ABIDE, OASIS-3, and ADNI, which differ in age range, clinical composition, longitudinal structure, and acquisition setting. We use chronological age as a common target because it is one of the few continuous variables available across all six datasets and is a standard reference task in connectome prediction~\cite{cole2017predicting}. Age is used here as a benchmark target rather than as a complete measure of biological aging, mental health, or clinical risk. Brain age analyses are known to be affected by regression to the mean and by residual coupling between prediction error and chronological age~\cite{de2020commentary}, so we interpret age prediction primarily as a controlled test of external generalization across open neuroimaging cohorts.

Each rs-fMRI scan included in the analysis is modeled as a connectome represented by a regularized SPD correlation matrix. In this paper, this term refers to a scan-wise SPD functional connectivity matrix obtained by estimating a regularized covariance matrix from regional time series, normalizing it to a correlation matrix, and applying numerical stabilization. We write \(\mathcal{S}_{++}^{p}\) for the set of \(p \times p\) SPD matrices, where \(p\) is the number of atlas regions. This set is the SPD manifold used throughout the paper. A common Riemannian approach centers SPD matrices at a reference point, often the Riemannian, or Fr\'echet, mean, and maps them to the tangent space at the reference point. Because this tangent space is a Euclidean vector space with the same dimension as the SPD manifold, standard machine learning methods can then be applied there. The Fr\'echet mean is the matrix on the SPD manifold that minimizes the sum of squared Riemannian distances to the training regularized SPD correlation matrices.

We compare two representative SPD matrix learning strategies. Tangent-Space Ridge maps each connectome to a fixed geometric representation centered at the training-set Fr\'echet mean and uses Ridge regression only for the downstream prediction task~\cite{abraham2017deriving}. SPDNet~\cite{huang2017riemannian} learns transformations that preserve SPD structure before applying a matrix logarithm and a prediction head. These models are compared with a Dummy regressor and a direct vectorized correlation baseline, CorrVec. We also evaluate Riemannian harmonization~\cite{honnorat2024riemannian}, fitted within each split so that every data-dependent quantity is estimated from the training split only~\cite{marzi2024efficacy}.

The contribution of this work is threefold. First, we provide a reproducible rs-fMRI benchmark across six datasets using these connectomes to test external generalization. It assesses whether gains from SPD matrix learning remain when a full dataset is held out. Second, we define within-dataset and pooled subject-grouped K-fold cross-validation (GroupKFold) protocols, together with leave-one-dataset-out (LODO) evaluation, leakage control, common model settings, and shared analysis scripts. Third, we show that within-dataset performance gives an important but incomplete view of robustness. When one dataset is held out for testing, method differences become smaller, prediction error increases, and age range mismatch and cohort differences strongly shape performance. This is the central point of the benchmark: the key question is not only which SPD matrix learning method performs best, but which gains remain when the cohort changes. The goal is to provide a transparent benchmark for evaluating future SPD matrix learning methods under realistic rs-fMRI dataset shift.

\begin{figure}[!tbp]
  \centering
  \includegraphics[width=0.9\linewidth]{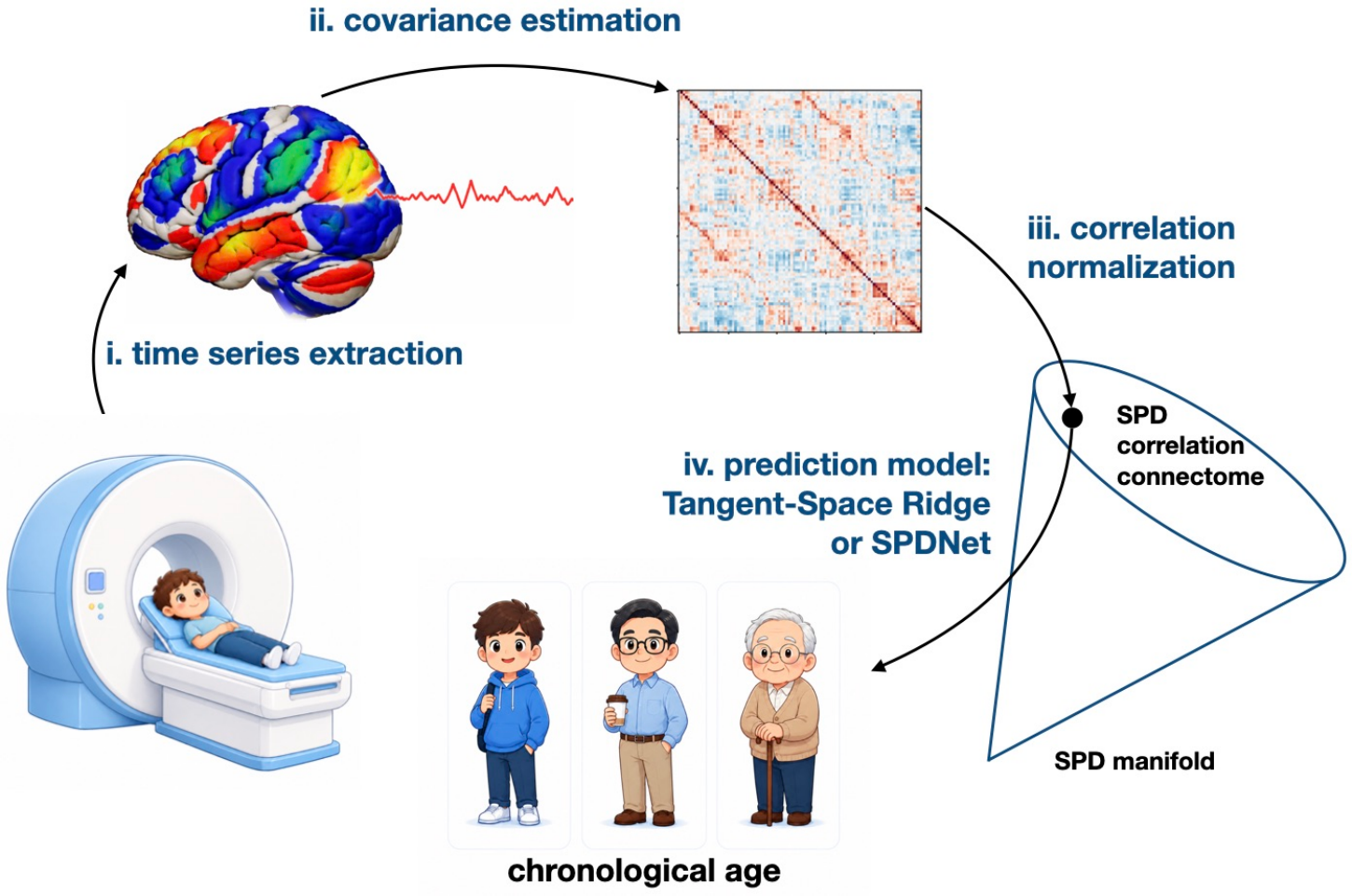}
    \caption{Overview of the rs-fMRI regression pipeline. Starting from each scan included in the analysis, we (i) extract regional time series, (ii) estimate a regularized covariance matrix, and (iii) normalize it to a correlation matrix. The resulting SPD correlation connectome is then passed to (iv) the prediction model, Tangent-Space Ridge or SPDNet, to predict chronological age.}
    \label{fig:main}
\end{figure}

\section{Methods}

The benchmark is designed to compare model generalization using a common connectome construction step after time series extraction. The overall predictive framework follows standard rs-fMRI connectome modeling practice~\cite{dadi2019benchmarking}. We implement it with open source Python tools, including scikit-learn~\cite{pedregosa2011scikit}\footnote{scikit-learn:\url{https://github.com/scikit-learn}}, Nilearn\footnote{Nilearn:\url{https://nilearn.github.io/stable/index.html}}, pyRiemann~\cite{pyriemann}\footnote{pyRiemann:\url{https://pyriemann.readthedocs.io/en/latest/\#}}, and SPD Learn~\cite{aristimunha2026spd}\footnote{SPD Learn:\url{https://github.com/spdlearn/spd_learn}}. The full pipeline is shown in Figure~\ref{fig:main}.

\subsection{Datasets}
\begin{figure*}[!t]
  \centering
  \includegraphics[width=\linewidth]{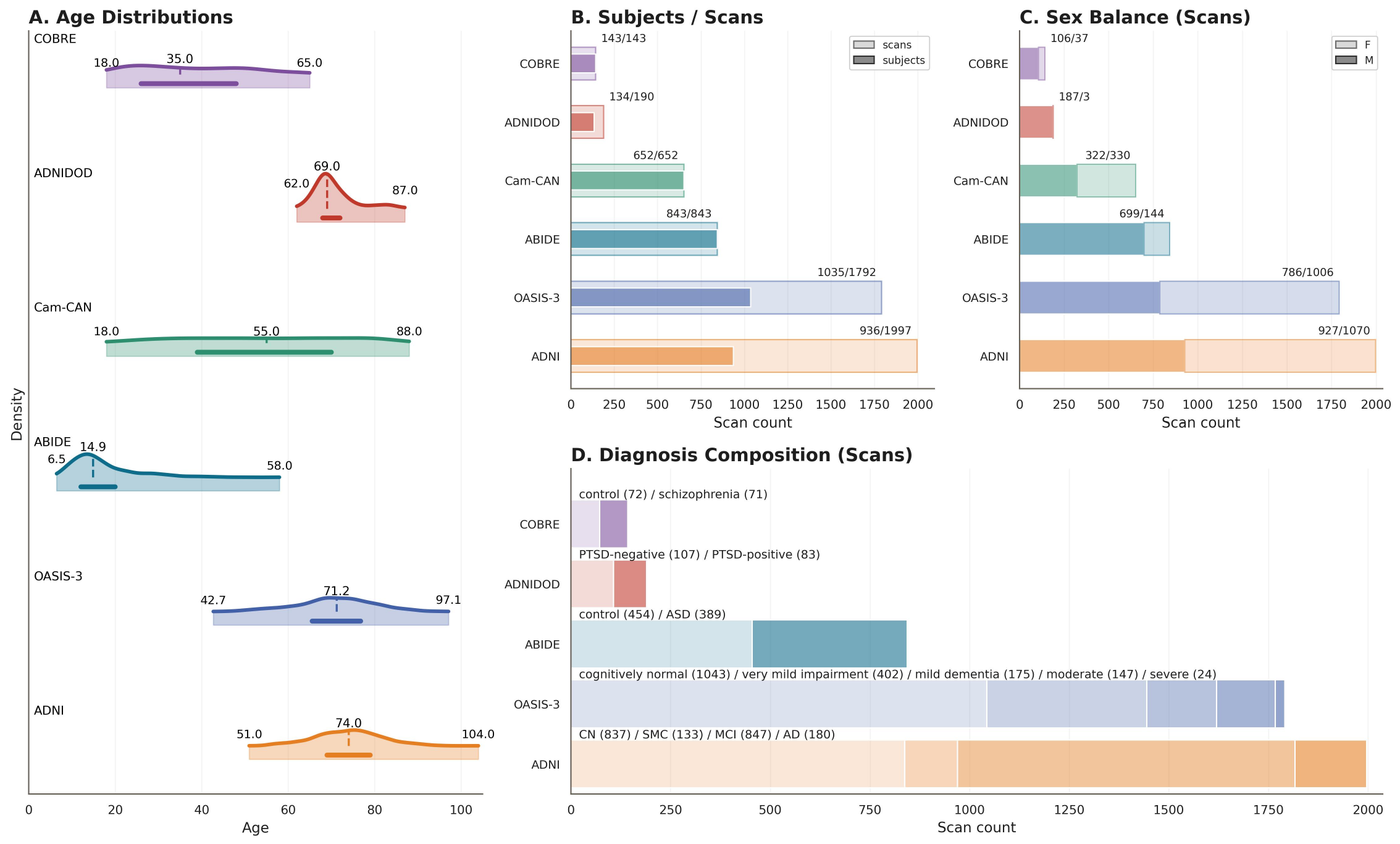}
  \caption{Overview of the six benchmark datasets (COBRE, ADNIDOD, Cam-CAN, ABIDE, OASIS-3, and ADNI) after the final inclusion and processing steps used in this study. Panel A shows the age range of each dataset: the left and right labels mark the minimum and maximum age, and the dashed line marks the median age. Panel B shows the numbers of subjects and scans. Panel C shows sex counts at the scan level. Panel D shows diagnosis composition for datasets where diagnosis labels are available in the processed tables. The figure highlights that the benchmark includes datasets with different age ranges, cohort sizes, and population compositions.}
  \label{fig:dataset_overview}
\end{figure*}

We used six publicly available or controlled access rs-fMRI datasets spanning psychiatric, developmental, healthy adult lifespan, and aging cohorts: COBRE~\cite{aine2017multimodal}, ADNIDOD~\cite{weiner2014effects}, Cam-CAN~\cite{shafto2014cambridge}, ABIDE~\cite{di2014autism}, OASIS-3~\cite{lamontagne2019oasis}, and ADNI~\cite{jack2008alzheimer}. After loading each dataset, matching phenotypes, extracting time series, and applying common quality control after extraction, the benchmark included 143 COBRE participants, 134 ADNIDOD participants with 190 scans, 652 Cam-CAN participants, 843 ABIDE participants, 1{,}035 OASIS-3 participants with 1{,}792 scans, and 936 ADNI participants with 1{,}997 scans. Figure~\ref{fig:dataset_overview} and Table~\ref{tab:dataset_context} summarize the final benchmark samples. Dataset-specific inclusion criteria and scan-to-phenotype matching procedures are provided in Supplementary Methods.

\subsection{Dataset Heterogeneity and Population Context}

Figure~\ref{fig:dataset_overview} and Table~\ref{tab:dataset_context} show that the six cohorts differ in age range and sample size, but also in population type, longitudinal structure, and site information. In brief, ABIDE is a mostly young autism dataset with multiple sites and the largest age mismatch relative to the rest of the benchmark. ADNIDOD, OASIS-3, and ADNI are older adult cohorts with repeated scans and diagnosis mixtures related to aging and dementia. Cam-CAN is the broadest healthy lifespan cohort. COBRE is a smaller psychiatric cohort covering young to middle adulthood.

These differences matter for interpretation. This benchmark is not only about site generalization, because the held-out folds can also differ in diagnosis mix, sex balance, and repeated measures structure. For example, the ABIDE cohort included in the analysis has a high proportion of male participants (699/843 scans), whereas the OASIS-3 and ADNI samples included in the analysis have more female than male participants. ADNI includes a mixed diagnosis spectrum spanning clinically normal (CN), subjective memory concerns (SMC), mild cognitive impairment (MCI), and mild Alzheimer's disease dementia (AD); COBRE includes schizophrenia and control participants at roughly equal frequency, and OASIS-3 includes several diagnosis categories rather than only healthy participants. Generalization across datasets in this study therefore combines demographic, clinical, longitudinal, and technical differences.

\begin{table*}[!tbp]
\caption{Summary of the six processed benchmark datasets. Diagnosis composition and final time series length are reported only when these fields are available in the processed analysis tables. For OASIS-3, CDR denotes Clinical Dementia Rating.}
\label{tab:dataset_context}
\centering
\small
\setlength{\tabcolsep}{3pt}
\begin{tabularx}{\textwidth}{@{}l >{\raggedright\arraybackslash\hsize=1.05\hsize}X >{\raggedright\arraybackslash\hsize=1.20\hsize}X >{\raggedright\arraybackslash\hsize=1.15\hsize}X >{\raggedleft\arraybackslash\hsize=0.60\hsize}X@{}}
\toprule
Dataset & Main cohort type & Diagnosis mix in scans included in the analysis & Scan structure / site info & Mean final time series length \\
\midrule
COBRE & Smaller psychiatric cohort from young to middle adulthood & Control 50.3\%, schizophrenia 49.7\% & One scan per subject with site information not available & 150.0 \\
ADNIDOD & Older adult cohort from ADNI-DOD & PTSD-negative 56.3\%, PTSD-positive 43.7\% & Repeated scans with 17 sites represented in the analysis & 156.3 \\
Cam-CAN & Healthy adult lifespan cohort & No diagnosis categories used & One scan per subject with site information not available & 256.0 \\
ABIDE & Mostly children, adolescents, and young adults & Control 53.9\%, ASD 46.1\% & One scan per subject with 19 sites represented in the analysis & 196.9 \\
OASIS-3 & Older adult aging / dementia cohort & CDR 0 58.2\%, 0.5 22.4\%, 1 9.8\%, 2 8.2\%, 3 1.3\% (one missing scan) & Repeated scans with site information not available & 159.0 \\
ADNI & Older adult aging / dementia cohort & CN 41.9\%, MCI 42.4\%, AD 9.0\%, SMC 6.7\% & Repeated scans with site information not available & 270.5 \\
\bottomrule
\end{tabularx}
\end{table*}

\subsection{Preprocessing and Denoising} 

We matched phenotypic variables to extracted fMRI time series and kept only scans with internally consistent identifiers and complete target information. Denoising could not be made identical across datasets because the available confound files differed across source releases. We therefore used nuisance regression appropriate for each dataset during time series extraction, while applying the same downstream connectome construction and evaluation logic to all datasets. Thus, results across datasets may reflect differences in cohorts, scanners, sites, and preprocessing, not only model performance.

Final quality control after extraction was applied uniformly using time series length, covariance conditioning, and the absence of regional time series columns with all zero values. The processed benchmark tables include final time series length and covariance quality checks, but they do not include scan-level motion measures or motion-based exclusion logs. Therefore, we could not measure residual motion effects in the same way across all datasets. The age prediction results should be interpreted as benchmark performance on processed functional-connectivity features, not as a pure measure of neural aging. Full denoising and quality control details are provided in Supplementary Methods.

\subsection{Time Series Extraction, Covariance Estimation, and Correlation Normalization}

We used the prespecified Schaefer-100 atlas for cortical parcellation~\cite{schaefer2018local}. Its 100-region resolution provides network-level detail while keeping connectome dimensionality manageable across datasets. Each scan was represented by a correlation-based SPD connectome. Connectomes were estimated scan-wise using the Oracle Approximating Shrinkage covariance estimator~\cite{chen2010shrinkage}. We added a small diagonal regularization term \(\epsilon I\), with \(\epsilon=10^{-5}\), to the covariance matrix and converted it to a correlation matrix by variance normalization. Note that variance normalization 
preserves positive definiteness of the regularized covariance matrix. In practice, we also added the same small diagonal jitter after correlation normalization as a numerical conditioning safeguard before downstream eigendecompositions and matrix-logarithm operations. The full construction recipe is provided in Supplementary Methods.

\subsection{Models}

We compare four prediction approaches: two SPD matrix learning methods and two reference baselines. The SPD methods are \emph{Tangent-Space Ridge}, a classical tangent-space method, and \emph{SPDNet}, a neural network method in SPD matrix learning. The reference baselines are Dummy and CorrVec.

We also evaluate their harmonized variants to test whether Riemannian harmonization changes performance. In these variants, harmonization is fitted using only the training data within each split before the same predictors are applied.

The tangent-space and harmonization formulas, SPDNet architecture, and the training, tuning, validation, and checkpointing settings are provided in Supplementary. 

Table~\ref{tab:comparison} summarizes the main differences between Tangent-Space Ridge and SPDNet in this benchmark.

\subsubsection{Tangent-Space Ridge}

Tangent-Space Ridge represents each SPD connectome by tangent-space coordinates at the training-set Fr\'echet mean under the affine-invariant Riemannian metric, and then fits a Ridge predictor~\cite{varoquaux2010detection}. 

Specifically, for each outer evaluation split, the Fr\'echet mean reference point is estimated from the training data only and then held fixed when transforming the corresponding evaluation data. 

The resulting tangent matrices are vectorized using the standard symmetric-matrix convention, with unit weights on diagonal entries and \(\sqrt{2}\) weights on off-diagonal entries. 

In both single-dataset and pooled benchmarks, tangent-space features are z-scored before Ridge regression. Feature means and standard deviations are estimated from the training split only and then applied unchanged to the corresponding validation or test data.

In the reported age benchmarks, Ridge regularization strength \(\alpha\) is selected within each outer training set using grouped inner cross-validation on the tangent features already computed from that outer training set.

\subsubsection{SPDNet}

SPDNet is a neural network architecture for learning from SPD matrices while maintaining their symmetric and positive-definite matrix structure through the learning process~\cite{huang2017riemannian}. In this study, it takes the full regularized SPD correlation connectome as input. The model applies learned bilinear matrix mappings and eigenvalue rectification, followed by a matrix logarithm and a multilayer regression head. After the matrix logarithm, the symmetric matrix is flattened using the upper triangle, including the diagonal, without the \(\sqrt{2}\) off-diagonal weighting used for Tangent-Space Ridge. 

The main SPDNet results use a fixed architecture and training protocol, with checkpoint selection based on a validation split that preserves subject groups within each outer training fold. 

For the SPDNet ablation in Figure~\ref{fig:result_4}, we compare four prespecified configurations that vary the BiMap output dimension and the number of BiMap--ReEig blocks.

\subsubsection{Dummy and CorrVec}

Dummy is an uninformative reference model that predicts the training-set mean age and ignores the connectome. 

CorrVec is a Euclidean functional-connectivity baseline that represents each correlation connectome by its off-diagonal upper triangular entries. The diagonal is excluded because correlation diagonals are constant. The retained off-diagonal entries are multiplied by \(\sqrt{2}\), giving an isometric upper-triangular vectorization of the off-diagonal symmetric matrix under the Frobenius inner product. CorrVec then uses the same grouped alpha selection and feature standardization as Tangent-Space Ridge.

\begin{table*}[!tbp]
\caption{Comparison of the two SPD matrix learning approaches used in this benchmark.}
\label{tab:comparison}
\centering
\small
\setlength{\tabcolsep}{4pt}
\begin{tabularx}{\textwidth}{>{\bfseries\raggedright\arraybackslash}p{0.15\textwidth}XX}
\toprule
\textbf{} & \textbf{Tangent-Space Ridge} & \textbf{SPDNet} \\
\midrule
\textbf{Input}
& A regularized SPD correlation connectome
& A regularized SPD correlation connectome \\
\midrule
\textbf{Models}
& Congruence whitening, a matrix logarithm at the identity matrix \(I\), and weighted upper-triangular vectorization
& BiMap, ReEig, and LogEig layers \\
\midrule
\textbf{Prediction head}
& Ridge regression on standardized tangent-space features
& A multilayer regression head after the matrix logarithm \\
\midrule 
\textbf{Reference and whitening}
& Uses the training-set Fr\'echet mean as a fixed reference point and applies congruence whitening before the log map at \(I\)
& Does not use a fixed Fr\'echet-mean reference point; BiMap layers learn congruence transformations before the matrix logarithm \\
\midrule
\textbf{Dimensionality handling}
& No learned dimension reduction in the reported benchmark
& BiMap layers can preserve or reduce the SPD matrix dimension \\
\midrule
\textbf{Harmonized variant}
& Uses harmonized tangent-space features
& Uses SPD matrices reconstructed after tangent-space harmonization \\
\bottomrule
\end{tabularx}
\end{table*}

\subsection{Harmonization Protocol}

To mitigate distribution shifts related to site while preserving the SPD matrix properties, we adopt \emph{Riemannian harmonization}~\cite{honnorat2024riemannian}. Each connectome is mapped to a tangent representation, corrected with a ComBat empirical Bayes model, and mapped back to the SPD manifold. In all reported harmonization analyses, dataset or site identity is modeled as the batch effect, and chronological age is included as the biological covariate to be preserved, whereas diagnosis, sex, motion, and scan quality variables are not included in the harmonization model. In this benchmark, harmonization is applied separately within each evaluation split so that information from held-out data is not used when fitting the harmonization model. Because chronological age is also the prediction target, results from harmonization with age as a covariate should be interpreted more narrowly.

\paragraph{Split-wise harmonization.}
In our implementation, we treat \texttt{SITE} as the batch effect and include \texttt{age} as a biological variable. To avoid leakage, harmonization is embedded within each outer training/test split~\cite{marzi2024efficacy}. Age is included because age distributions differ strongly across datasets, making age partly confounded with the batch variable. Without age in the covariate design, ComBat-style correction can attribute some age-related FC variation to site or dataset effects. Including age therefore preserves the modeled age effect while correcting batch terms, consistent with harmonization work that models lifespan age effects as biological variation to be preserved~\cite{pomponio2020harmonization}.

When cross-validation is performed on the pooled sample, test-fold harmonization uses true test ages. These results should therefore be read as harmonization with age as a covariate, not as age prediction in a setting where test age is unavailable. When a LODO scheme is used, harmonization is restricted to the pooled training datasets; the held-out dataset is not harmonized using its true ages.

Throughout the pooled analyses, a superscript $*$ denotes the split-wise harmonized variant of a method. Thus, $*$ denotes test-fold harmonization in pooled GroupKFold and training-side harmonization in LODO. Tangent-Space Ridge$^{*}$ fits Ridge regression on the ComBat-corrected tangent features. CorrVec$^{*}$ and SPDNet$^{*}$ first reconstruct harmonized SPD matrices with the exponential map at the split-specific training reference point; CorrVec then vectorizes these matrices, whereas SPDNet uses them as SPD inputs.

\subsection{Evaluation Protocols}

We conduct two types of experiments. First, we perform GroupKFold cross-validation separately on each dataset. Second, we run both GroupKFold and LODO evaluations on the union of the six datasets. In all GroupKFold settings, we set $K=5$. Within-dataset performance is summarized with negative MAE and the coefficient of determination ($R^2$). For the pooled benchmark, we additionally report the LODO Spearman rank correlation coefficient (Spearman $\rho$) between predicted and true age. All experiments were run with fixed random seeds for reproducibility. The reported results come from this fixed random seed setup and should not be taken as evidence that the same ranking would hold across other random seeds.

The outer GroupKFold protocol groups scans by subject so that the same subject cannot appear in both train and test within a fold. In the pooled benchmark across six datasets, subject identifiers are prefixed with the dataset label before grouping, which prevents collisions between nominally identical subject IDs across datasets. Within each outer training fold, 10\% of the outer training subjects or subject groups are reserved for validation using a random split that respects subject groups and a fixed seed. In LODO, the outer test split is defined by leaving out one entire dataset, whereas the internal train/validation split still uses the same grouped procedure on the pooled training datasets.

\subsection{Evaluation Metrics}

Because the benchmark uses age prediction as a regression task, we summarize performance with complementary metrics. For all regression settings, we report \emph{negative mean absolute error} (NegMAE) and the coefficient of determination $R^2$. NegMAE is the negative of the mean absolute prediction error, so larger values indicate lower error and the scale remains interpretable in age units. The coefficient of determination $R^2$ measures how much variance in chronological age is explained relative to a baseline that predicts the mean. Within the same data distribution, NegMAE and $R^2$ together provide a compact summary of accuracy and explained variance.

For LODO, the held-out dataset may have a very different age distribution from the pooled training datasets, so error metrics alone can be difficult to interpret. We therefore also report \emph{Spearman $\rho$}, the rank correlation between predicted and true age within the held-out dataset. Spearman $\rho$ measures whether the model preserves age ordering even when absolute predictions are poorly calibrated. Together, NegMAE, $R^2$, and Spearman $\rho$ separate prediction error, explained variance, and rank-order preservation under held-out-cohort evaluation.

\section{Results}

\subsection{\texorpdfstring{Dataset Differences Define the External Generalization Problem}{Dataset Differences Define the External Generalization Problem}}

The six benchmark datasets are not minor variations of the same population. Figure~\ref{fig:dataset_overview} and Table~\ref{tab:dataset_context} show clear differences in age range, cohort size, diagnosis mix, sex balance, longitudinal structure, and site information. ABIDE participants are much younger than the rest of the benchmark, Cam-CAN spans the broadest healthy adult lifespan, COBRE is a smaller psychiatric cohort, and ADNIDOD/OASIS-3/ADNI are older adult cohorts with repeated scans and mixed diagnosis structure. These differences define the external generalization problem. They also mean that transfer across datasets in this study should not be read as scanner/site effects alone, but as a combination of demographic, clinical, longitudinal, and technical differences.

\subsection{\texorpdfstring{Within-Dataset Accuracy Is Useful but Incomplete}{Within-Dataset Accuracy Is Useful but Incomplete}}

\begin{figure}[!htbp]
  \centering
  \includegraphics[width=\linewidth]{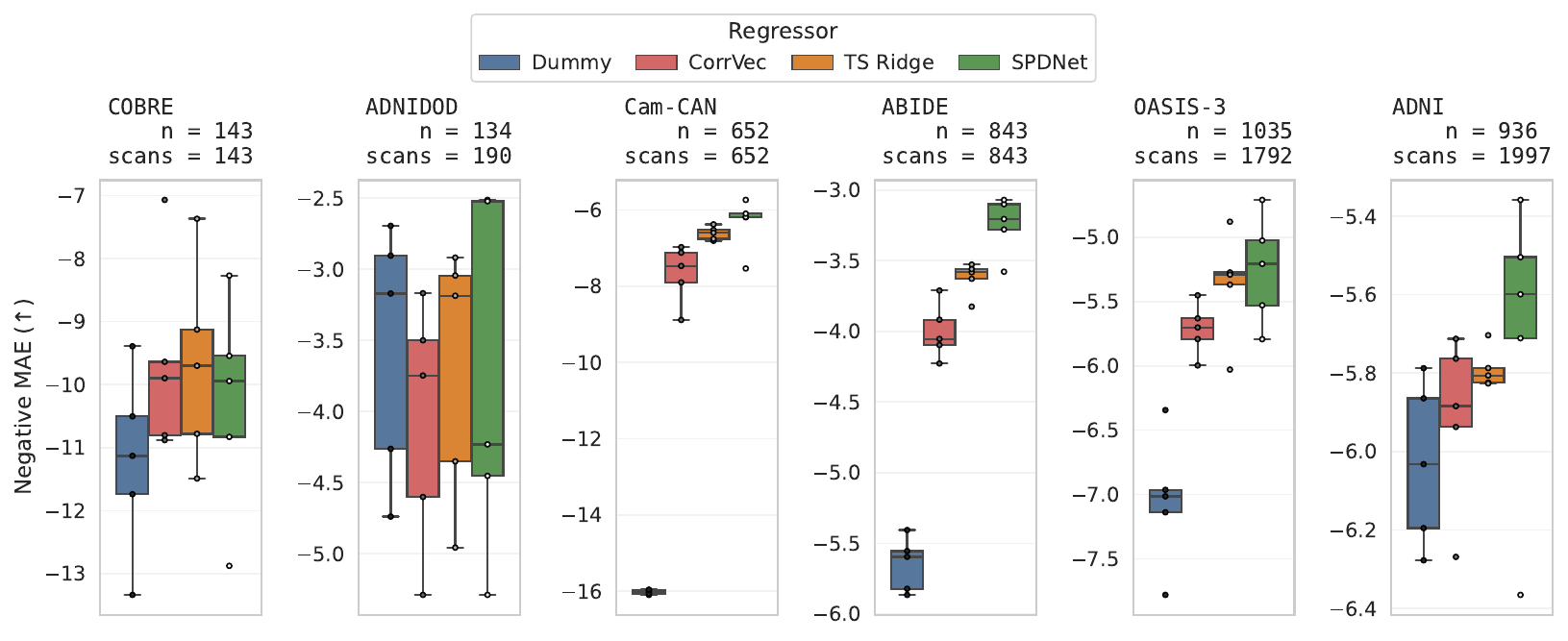}
  \includegraphics[width=\linewidth]{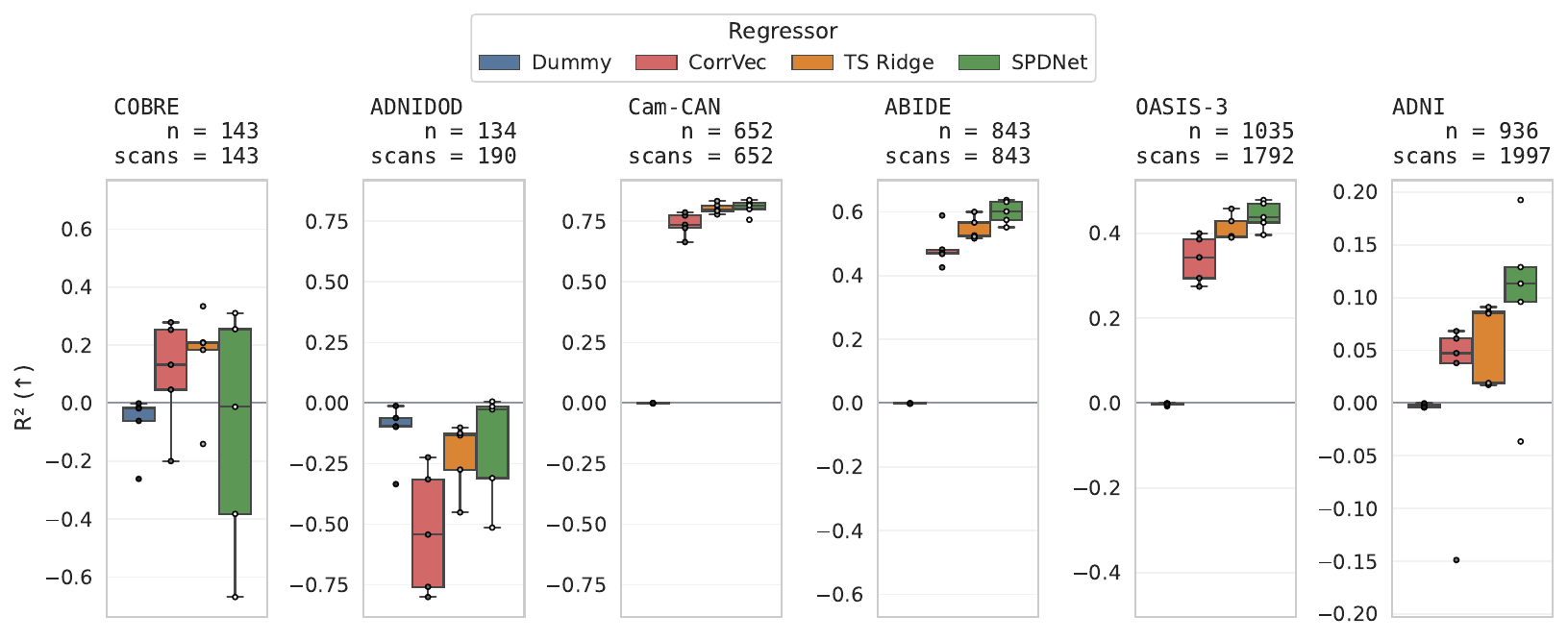}
    \caption{\textbf{Within-dataset prediction benchmark} across COBRE, ADNIDOD, Cam-CAN, ABIDE, OASIS-3, and ADNI. The datasets are arranged from left to right from fewer to more scans included in the analysis, and the annotations above each column report the number of participants and scans included in the analysis. For each dataset, the top panel reports GroupKFold negative MAE and the bottom panel reports GroupKFold $R^2$. Boxplots compare Dummy, CorrVec, Tangent-Space Ridge, and SPDNet.}
    \label{fig:result_1}
\end{figure}

Figure~\ref{fig:result_1} summarizes GroupKFold performance within each dataset. The figure reports both negative MAE and $R^2$ for Dummy, CorrVec, Tangent-Space Ridge, and SPDNet, with sample sizes and scan counts shown above each dataset panel. From left to right, the datasets are ordered from fewer to more scans included in the analysis. Except for the smaller and more variable cohort, ADNIDOD ($n=134$, 190 scans), all informative models outperform the Dummy baseline on most folds. This shows that the connectomes retain age-related signal within datasets.

Across datasets, CorrVec is a useful Euclidean baseline, but Tangent-Space Ridge and SPDNet are usually as good or better in both metrics. The best within-dataset performance appears in Cam-CAN, ABIDE, OASIS-3, and ADNI, where all informative models reach positive $R^2$ and SPDNet often has the best median negative MAE. On COBRE, the gap between CorrVec, Tangent-Space Ridge, and SPDNet is smaller. On ADNIDOD, the distributions remain broad and $R^2$ stays near or below zero for all informative methods. These within-dataset results are useful, but they should not be read as a direct measure of robustness across datasets. In pooled settings, dataset identity, age structure, and dataset composition may still provide indirect shortcuts that are absent under LODO.

\subsection{\texorpdfstring{Pooled Evaluation Gives a More Optimistic View Than LODO}{Pooled Evaluation Gives a More Optimistic View Than LODO}}

Figure~\ref{fig:result_2} summarizes the pooled benchmark across six datasets under GroupKFold and LODO. The figure reports GroupKFold and LODO negative MAE, GroupKFold and LODO $R^2$, and LODO Spearman $\rho$. We compare three method families, CorrVec, Tangent-Space Ridge, and SPDNet, each in original form and after harmonization (CorrVec$^{*}$, Tangent-Space Ridge$^{*}$, and SPDNet$^{*}$). The Dummy regressor is kept only as a reference summary in the panel annotations.

Under pooled GroupKFold, all informative models perform better than the Dummy reference by a wide margin. CorrVec$^{*}$ improves over CorrVec, Tangent-Space Ridge$^{*}$ shows an even larger gain over Tangent-Space Ridge, and SPDNet has the best overall negative MAE and $R^2$ in this pooled setting, with only a small difference between SPDNet and SPDNet$^{*}$. These pooled GroupKFold scores should be interpreted carefully. Because every dataset appears in both training and testing, strong performance can reflect relationships between the connectomes and age at the scan level, but it can also reflect broader cohort structure that is shared across folds.

Under LODO, performance drops relative to pooled GroupKFold. Prediction error increases, $R^2$ becomes unstable and is often below zero, and the spread across held-out datasets increases for all methods. One LODO point is consistently more extreme than the others. This point corresponds to the ABIDE fold and is not a plotting artifact. ABIDE is much younger ($17.13 \pm 7.62$ years, median $14.86$) than the other five datasets, whose pooled age distribution is centered in adulthood and older age (Figure~\ref{fig:dataset_overview}), so the held-out ABIDE point appears far from the other points in MAE and $R^2$. Spearman $\rho$ is not reported for the Dummy regressor because Dummy gives constant predictions within each held-out dataset, making rank correlation undefined. Even when absolute errors are large and $R^2$ is poor, Spearman $\rho$ remains positive for several held-out datasets, especially for Tangent-Space Ridge and SPDNet. This indicates that some age rank-order information can remain under external dataset shift even when calibration and absolute accuracy degrade. Harmonization helps CorrVec and Tangent-Space Ridge more consistently than SPDNet, while SPDNet keeps competitive median error but shows wider variability across datasets.

This drop in performance should be interpreted in the context of more than site effects. The held-out cohorts also differ in population composition: COBRE and ABIDE are psychiatric datasets, OASIS-3 and ADNI include aging and dementia samples, and ADNIDOD/OASIS-3/ADNI are partly longitudinal (Table~\ref{tab:dataset_context}). As a result, LODO combines site and acquisition shift with demographic and clinical differences between datasets.

Figure~\ref{fig:dataset_overview} also helps explain two secondary patterns. Because ADNIDOD covers a narrow and relatively old age range, $R^2$ and Spearman $\rho$ should be interpreted cautiously for this held-out fold. By contrast, Cam-CAN spans a broad adult lifespan and overlaps more strongly with the other datasets, which helps explain why its LODO fold is often among the most stable.

Table~\ref{tab:lodo_per_dataset} makes these differences across folds explicit. SPDNet gives the best MAE and $R^2$ on ABIDE, OASIS-3, and ADNI, whereas COBRE and Cam-CAN are more competitive for shallower models depending on the metric. At the same time, the best Spearman $\rho$ does not always come from the same model that minimizes MAE, reinforcing the value of separating absolute error from rank-order information under LODO.

\subsection{\texorpdfstring{Harmonization Benefits Depend on the Evaluation Protocol}{Harmonization Benefits Depend on the Evaluation Protocol}}

\begin{figure}[!tbp]
  \centering
  \includegraphics[width=\linewidth]{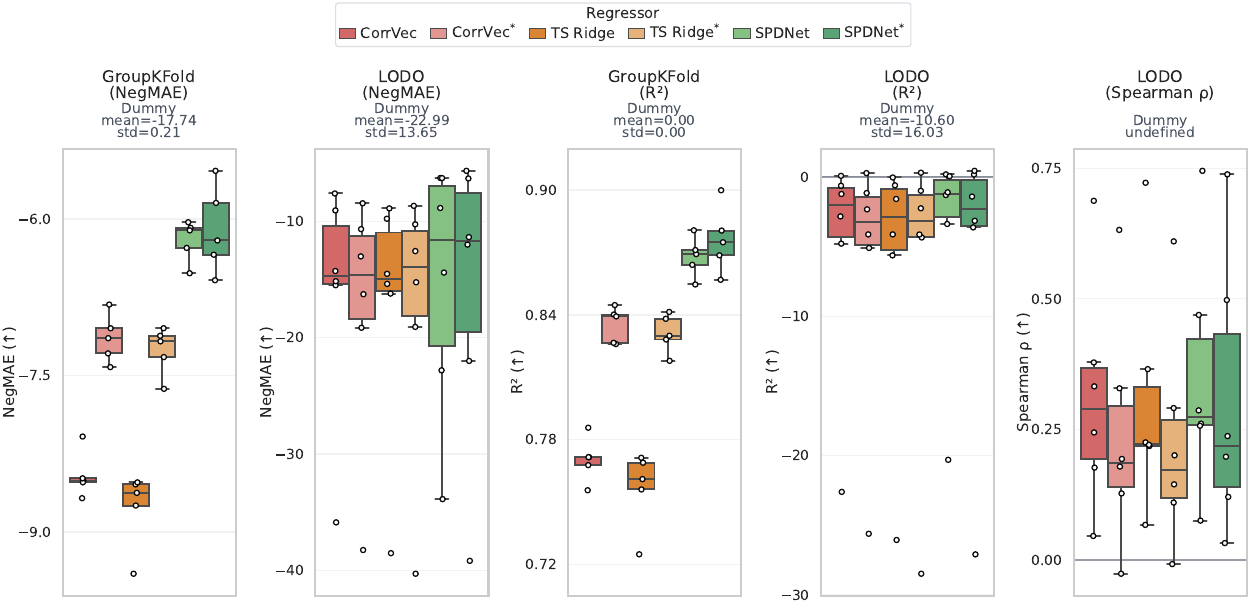}
    \caption{
    \textbf{Across-dataset prediction benchmark} on the pooled sample from COBRE, ADNIDOD, Cam-CAN, ABIDE, OASIS-3, and ADNI. The five panels report GroupKFold negative MAE, LODO negative MAE, GroupKFold $R^2$, LODO $R^2$, and LODO Spearman $\rho$. Boxplots compare CorrVec, Tangent-Space Ridge (TS Ridge), and SPDNet, each in original form and after harmonization. A superscript $*$ denotes the harmonized version of a method. Dummy scores are shown as panel annotations because the Dummy regressor is a constant-prediction reference; its Spearman $\rho$ is undefined. Each LODO point corresponds to one held-out dataset.}
    \label{fig:result_2}
\end{figure}

Figure~\ref{fig:result_2} also shows that harmonization can help, but its interpretation depends on the evaluation protocol. CorrVec$^{*}$ and Tangent-Space Ridge$^{*}$ improve clearly over their original versions in pooled GroupKFold, and Tangent-Space Ridge$^{*}$ keeps a smaller but still visible advantage in LODO. These gains are useful benchmark findings, but they depend on the protocol and should not be read as a general guarantee of transfer across datasets. In pooled GroupKFold, the harmonized results use true test age during feature correction. In LODO, the held-out dataset remains untouched, so the harmonized results reflect preparation of the training data only. The benchmark therefore supports a narrower lesson: harmonization can help, but its measured benefit depends on both the evaluation protocol and whether age is used during feature correction.

\begin{table*}[!tbp]
\caption{LODO summary for each held-out dataset. Each entry reports the best performing model or setting for MAE, $R^2$, and Spearman $\rho$, based on the exported LODO metric summary. Ridge denotes Tangent-Space Ridge. A superscript $*$ denotes the harmonized version of a model. Methods without $*$ use the original features.}
\label{tab:lodo_per_dataset}
\centering
\small
\renewcommand{\arraystretch}{1.08}
\setlength{\tabcolsep}{0pt}
\begin{tabular}{@{}l@{\hspace{1.6em}}l@{\hspace{1.6em}}l@{\hspace{1.6em}}l@{}}
\toprule
Held-out dataset & Lowest MAE & Highest $R^2$ & Highest Spearman $\rho$ \\
\midrule
COBRE & CorrVec, 15.15 & CorrVec, -1.21 & CorrVec, 0.332 \\
ADNIDOD & SPDNet$^{*}$, 5.66 & SPDNet, -1.08 & SPDNet, 0.075 \\
Cam-CAN & SPDNet$^{*}$, 11.36 & SPDNet$^{*}$, 0.460 & SPDNet, 0.745 \\
ABIDE & SPDNet, 33.88 & SPDNet, -20.30 & SPDNet, 0.261 \\
OASIS-3 & SPDNet, 6.27 & SPDNet, 0.204 & SPDNet$^{*}$, 0.498 \\
ADNI & SPDNet, 8.86 & SPDNet, -1.28 & SPDNet, 0.286 \\
\bottomrule
\end{tabular}
\end{table*}

\subsection{\texorpdfstring{Larger SPDNet Configurations Do Not Consistently Improve Transfer}{Larger SPDNet Configurations Do Not Consistently Improve Transfer}}

\begin{figure}[!tbp]
  \centering
  \includegraphics[width=\linewidth]{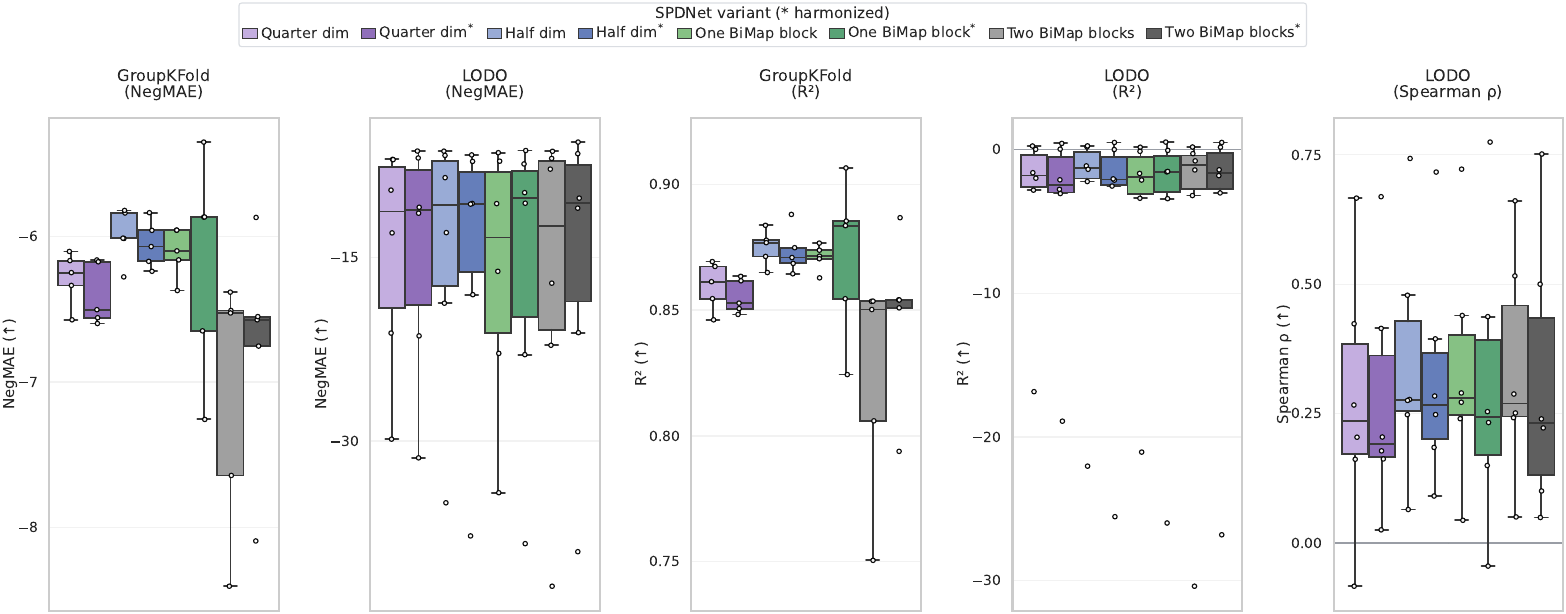}
    \caption{SPDNet configuration ablation under pooled GroupKFold and LODO evaluation. The five panels report GroupKFold negative MAE, LODO negative MAE, GroupKFold $R^2$, LODO $R^2$, and LODO Spearman $\rho$. Boxplots compare four SPDNet variants: quarter dimension, half dimension, one BiMap block, and two BiMap blocks, each in original form and after harmonization. A superscript $*$ denotes the harmonized version of a variant. Overall, no single SPDNet configuration is consistently best across these diagnostics.}
    \label{fig:result_4}
\end{figure}

Figure~\ref{fig:result_4} compares four SPDNet configurations for age prediction: quarterdim (BiMap+ReEig), halfdim (BiMap+ReEig), one (BiMap+ReEig), and two repeated BiMap+ReEig blocks, each in original form and after harmonization. The figure reports the same diagnostics for the pooled benchmark as Figure~\ref{fig:result_2}: GroupKFold and LODO negative MAE, GroupKFold and LODO $R^2$, and LODO Spearman $\rho$. Overall, the figure does not support one best SPDNet configuration. The smaller quarterdim and halfdim variants are competitive under pooled GroupKFold, while the larger one and two variants do not show a reliable advantage across both pooled GroupKFold and LODO settings. Under LODO, all configurations are less stable than under GroupKFold, and no configuration is consistently best in error, explained variance, or rank preservation. Harmonization changes some medians slightly, but it does not produce a stable ranking across these diagnostics. Because all plotted values come from the fixed random seed benchmark protocol described above, this ablation should be read as a controlled configuration comparison at fixed seeds, not as evidence of robustness across different random seeds.

\section{Discussion}

Viewed as a benchmark of external generalization, the results have four main lessons for rs-fMRI prediction studies. First, connectomes built with the shared construction pipeline contain age-related signal within individual datasets. Most informative models outperform the Dummy baseline in within-dataset GroupKFold cross-validation, especially in larger cohorts with broader age ranges (Figure~\ref{fig:result_1}). At the same time, performance varies across datasets, which shows that model performance cannot be judged apart from sample size, age range, and cohort composition. This matters for neuroimaging because individual prediction is often used as evidence that FC carries age-related signal. Without external validation, that evidence can be partly driven by cohort structure and benchmark design.

Second, the model comparison does not support a simple hierarchy in which the deep SPD matrix learning model always beats the shallower alternatives. CorrVec is already a strong Euclidean reference. Tangent-Space Ridge is competitive on several smaller or more variable datasets. SPDNet gives modest gains in some larger datasets, but these gains are not consistent enough to support a general ranking across settings. This supports SPDNet as an important benchmark method while preserving the need for simple baselines.

Third, the main pattern in the study is not the difference between SPD matrix learning models, but the difference between evaluation within the training distribution and external validation. When cross-validation (GroupKFold) is on pooled data, every dataset appears in both training and testing, and all informative models outperform Dummy by a wide margin. Under LODO cross-validation, the model must generalize to a held-out dataset. In this setting, method differences narrow, prediction error increases, and $R^2$ often becomes negative (Figure~\ref{fig:result_2}). Spearman $\rho$ adds useful information because it shows whether age rank-ordering is preserved within the held-out dataset even when absolute error is large or $R^2$ is poor. This pattern supports the use of multiple diagnostics rather than a single error metric when testing neuroimaging predictors under dataset shift.

Fourth, harmonization is useful, but its meaning depends on the evaluation protocol. Tangent-Space Ridge benefits clearly from Riemannian harmonization in pooled GroupKFold and to a smaller extent in LODO, whereas SPDNet shows smaller and less stable gains. This harmonization step follows the Riemannian harmonization framework of Honnorat et al.~\cite{honnorat2024riemannian}; the present benchmark evaluates how such split-wise harmonization behaves under pooled GroupKFold and strict LODO transfer. The larger gains for Tangent-Space Ridge indicate that the effect of harmonization depends on the downstream model, although this benchmark does not identify the mechanism. More importantly, the estimated benefit of harmonization depends on the protocol, especially on whether age is used during feature correction.

\subsection{Dataset Differences Beyond Site Effects}

An important point from Table~\ref{tab:dataset_context} is that generalization across datasets in this benchmark should not be read as scanner/site effects alone. The datasets included in the benchmark differ in age range, diagnosis mix, sex balance, and longitudinal structure. ABIDE is a younger autism cohort with a high proportion of male participants and 19 sites represented in the analysis. COBRE is a smaller schizophrenia/control cohort. Cam-CAN is the closest to a healthy lifespan reference cohort. ADNIDOD, OASIS-3, and ADNI are older adult cohorts with mixed diagnosis composition and many repeated scans. These differences make the benchmark more realistic, but they also mean that lower performance on a held-out dataset can reflect population and technical differences.

This matters especially for the LODO results. When ABIDE is held out, the main mismatch is age and developmental stage. When COBRE is held out, psychiatric group differences are mixed with age and limits from sample size. When ADNI or OASIS-3 is held out, the model may also face longitudinal aging and dementia structure, which is not the same as healthy aging. Table~\ref{tab:lodo_per_dataset} shows this pattern at the fold level: ABIDE remains difficult even for the best model, Cam-CAN is relatively stable, and the best model changes across datasets and metrics. In other words, the benchmark does not point to one single source of difficulty. Different held-out datasets expose different combinations of age-range extrapolation, age-related bias, and mismatch between datasets. This benchmark is therefore not a pure study of domain adaptation. It measures robustness to joint site, acquisition, and population differences across open rs-fMRI cohorts.

\subsection{Interpreting Harmonization and Model Performance}

The LODO results suggest that harmonization is helpful only for some forms of dataset shift. In this benchmark, harmonization is applied to the full tangent space feature set using \texttt{SITE} and \texttt{age} as variables, with empirical Bayes enabled and age modeled as a smooth term bounded to $(0,120)$. This setup uses age during harmonization. It is useful for studying harmonization, but it is not the same as a prediction setting where test age is unknown. The correction depends on separating site effects from age effects, which is difficult when datasets have different age ranges.

These implementation choices can affect the observed results. Empirical Bayes shrinkage is expected to improve stability across datasets that differ from each other and across high dimensional tangent features, but it may also make the correction more conservative. The smooth age term helps model nonlinear age effects, but it does not solve the more basic problem that some LODO folds have weak overlap in age range with the training data. This is especially important for ABIDE (Figure~\ref{fig:dataset_overview}), whose much younger age distribution makes the LODO fold closer to prediction outside the training age range than to ordinary transfer across datasets.

The setup difference between GroupKFold and LODO is also important. In pooled GroupKFold, harmonization is applied to both train and test folds, and the test fold transformation uses the true \texttt{age} variable. In LODO, by contrast, harmonization is fitted on the training datasets only and the held-out dataset is left uncorrected. Thus, pooled GroupKFold evaluates harmonization when test ages are known, while LODO applies harmonization only to the training datasets and leaves the held-out dataset unchanged. More generally, the findings should be read as properties of this tangent space harmonization pipeline, including the numerical cleanup applied after harmonization, rather than as a statement about harmonization that is independent of representation choice. Overall, harmonization appears most useful when technical differences are the main source of shift, and less useful when the main problem is a large demographic mismatch between training and test cohorts. A stronger test would compare ComBat without age with ComBat that includes age, but that comparison is not included in the present benchmark.

The SPDNet ablation in Figure~\ref{fig:result_4} compares four configurations using the same fixed random seed setup. It is not a full search over SPDNet architectures. Once multiple configurations are compared together, the benchmark does not show that deeper or larger architectures are better. The quarterdim, halfdim, one block, and two block variants have broadly overlapping ranges, and their ordering changes across GroupKFold error, LODO error, $R^2$, and LODO Spearman $\rho$. The safest interpretation is empirical: in the present data regime, changing SPDNet width or depth within this family does not reliably produce one configuration that is best across datasets, error, explained variance, and rank preservation. This could reflect limited sample size, harder optimization, or partial redundancy between architectural flexibility and the already standardized input pipeline. The benchmark only shows that no single SPDNet configuration is consistently best in this experiment.

These observations also clarify how age prediction should be interpreted in this study. Because chronological age is available across all six datasets, it provides a common regression target for comparing methods under a single evaluation framework. This makes it useful for benchmarking robustness and transfer. However, the results should not be interpreted as a direct measure of biological aging or clinical utility. In particular, MAE and $R^2$ across datasets remain sensitive to changes in age distribution across datasets, so the strongest claim supported by the present experiments is methodological: they quantify how well different SPD learning pipelines retain predictive signal under dataset shift.

Taken together, the results support a focused conclusion. The main challenge in this benchmark is not extracting age-related connectome signal within a dataset, but preserving that signal under large shifts across datasets. Age range mismatch, cohort heterogeneity, and remaining age-related bias are all important contributors, whereas harmonization provides a clearer benefit for the fixed tangent space pipeline than for SPDNet.

\subsection{Toward More Rigorous External Evaluation}

This benchmark suggests four practical points for future work. First, studies should report LODO or other external validation across datasets in addition to within-dataset cross-validation, because within-dataset scores are useful but incomplete as evidence of external robustness. These results do not invalidate earlier within-dataset results, but they suggest that such results should be interpreted together with external validation tests when the goal is application to new cohorts. Second, papers should report overlap in age range and other main dataset differences before interpreting performance changes across datasets. Third, simple baselines such as CorrVec and Tangent-Space Ridge should remain in the comparison set because they provide strong reference points that are easy to read. Fourth, benchmark releases should publish the train/test splits, model settings, exported metric tables, and figure scripts so that new models for SPD correlation connectomes can be tested under the same protocol. These practices would make rs-fMRI individual prediction results easier to compare across laboratories and datasets.

This distinction is also relevant to recent fMRI foundation models, although such models are not evaluated in this benchmark. BrainLM and NeuroSTORM report promising transfer across cohorts, tasks, or downstream settings~\cite{ortega2024brainlm,wang2026towards}. However, external adaptation and strict external validation answer different questions. Fine-tuning or linear probing on a target cohort tests whether a learned representation remains useful after target cohort labels are available. The LODO setting used here is stricter in one specific sense: the held-out cohort contributes no samples to model fitting, harmonization, scaling, validation, or model selection, and is used only for testing. This setting remains important because practical application often begins with applying a trained pipeline to a new cohort before enough labels are available for adaptation. Thus, even in the foundation model era, external cohort generalization remains an open evaluation problem, and future large fMRI models would benefit from reporting LODO tests together with strong classical connectome baselines.

More broadly, these results suggest that external cohort validation should be treated as a core part of rs-fMRI prediction studies, not as an optional robustness check. A model that performs well within one dataset has answered an important question, but not yet the question that matters most for application beyond the development cohort: whether the same pipeline can be carried to a new cohort without using that cohort during model development.

The intended use of this benchmark is therefore simple. When a new rs-fMRI prediction model claims broad generalization, it should be compared with strong connectome baselines under a held-out cohort test. Reporting only within-dataset performance leaves the most important question unanswered. The value of the benchmark is to make that question concrete and reproducible.

\section{Conclusions}

We present a reproducible benchmark across six datasets for testing SPD matrix learning methods under rs-fMRI dataset shift. The benchmark uses shared connectome construction after extraction, splitting by subject, leakage control within each split, and common model settings. It compares CorrVec, Tangent-Space Ridge, SPDNet, and Riemannian harmonization on these connectomes, using chronological age as a common target across datasets. Its main purpose is to evaluate whether rs-fMRI individual prediction pipelines can be applied across cohorts, rather than to rank SPD models in isolation.

The main finding is that within-dataset and pooled evaluations give an important but incomplete view of external robustness. Informative models detect age-related connectome signal within datasets, but their advantages become smaller under LODO. The performance drops under LODO are not explained by scanner/site effects alone. They reflect a mixture of age range mismatch, cohort composition, longitudinal structure, diagnosis mix, and preprocessing differences. This is why rank-order correlation is a useful complement to MAE and $R^2$ in external validation across datasets.

The study also clarifies the interpretation of the two SPD matrix learning modeling strategies. Tangent-Space Ridge provides a strong and transparent geometric baseline. SPDNet is competitive and sometimes stronger within datasets, but larger SPDNet variants do not consistently improve transfer in the fixed seed benchmark. Riemannian harmonization helps most for tangent space features, while its benefit for SPDNet is smaller and more variable. These results support a conservative conclusion: SPD geometry is useful for organizing the benchmark, but external generalization is still strongly shaped by dataset shift.

Several limitations should be kept in mind. Age is a practical shared target, not a complete biological or clinical outcome. Pooled GroupKFold may use dataset identity information that is absent in strict external validation, and a baseline relying only on metadata would help quantify this effect. Pooled harmonization uses true test age when correcting the test fold, so it does not represent a prediction setting where test age is unknown. The LODO benchmark also mixes technical differences with age range mismatch, especially for ABIDE. Further analyses of age range overlap, motion sensitivity analyses, GSR sensitivity analyses, broader runs across random seeds, and nonlinear Euclidean baselines would strengthen the comparison. Finally, scan-level motion measures and motion-based exclusion logs were not available across all datasets. Residual motion may therefore still contribute to the functional-connectivity features and may vary with age or dataset membership. For this reason, the present results should not be interpreted as isolating neural aging effects from motion-related effects. Despite these limits, the benchmark provides a transparent reference point for evaluating future SPD matrix learning methods under realistic rs-fMRI dataset shift.

\markboth{DECLARATIONS}{DECLARATIONS}
\section*{Data and Code Availability}

All six datasets used in this study are public community resources and must be obtained from their original providers under the corresponding data use agreements. This benchmark does not redistribute any raw neuroimaging data. In accordance with the licensing and redistribution policies of the source datasets, only redistribution-permitted derived functional connectivity matrices from CamCAN, ABIDE, and COBRE are publicly released through Zenodo (version 0.1.0; \url{https://doi.org/10.5281/zenodo.21912219})~\cite{ju2026benchmarkdata}. No data derived from ADNI, ADNI-DOD, or OASIS-3 are redistributed; users wishing to reproduce the corresponding experiments must obtain these datasets directly from their original providers. The Zenodo release also contains the released derived benchmark files, including \texttt{.npz} files with connectome inputs, derived metadata, and split definitions for the redistribution-permitted datasets. Benchmark splits are reconstructed from subject and dataset identifiers using the documented protocols with fixed random seeds. This design enables other researchers to reproduce the benchmark, inspect the evaluation protocol, and compare new SPD matrix learning methods under identical experimental settings. The analysis code, model configuration files, split logic, and scripts are available at \url{https://github.com/GeometricBCI/rsfmri-spd-connectome-external-generalization-benchmark}.

\section*{Author Contributions}

Ce Ju conceived the study, designed the benchmark, implemented the analyses, ran the experiments, analyzed the results, prepared the figures and tables, organized the benchmark release, and wrote the manuscript. Antoine Collas provided the data processing files used in the benchmark preparation and contributed to early exploratory analyses. Florent Bouchard carefully reviewed the manuscript and provided insightful comments on its mathematical aspects. Bertrand Thirion supervised the study, provided scientific guidance, contributed to interpretation of the results, and reviewed and edited the manuscript. All authors approved the final manuscript.

\section*{Funding}

This work was supported by grant ANR-22-PESN-0012 under the France 2030 program, managed by the Agence Nationale de la Recherche (ANR), and by the DATAIA Convergence Institute under the ``Programme d'Investissement d'Avenir'' (ANR-17-CONV-0003), operated by Inria.

\section*{Declaration of Competing Interests}

The authors declare no competing interests.

\section*{Ethics Statement}

This study reanalyzes deidentified data from public or controlled access neuroimaging resources. No new human participant data were collected for this benchmark. Ethical approval and informed consent procedures were handled by the original studies and data providers.

\section*{Use of AI-Assisted Tools}

OpenAI Codex was used for English-language editing and to organize the submission files. The authors reviewed the resulting text and take full responsibility for the scientific content and final manuscript.

\clearpage
\printbibliography

\clearpage
\markboth{SUPPLEMENTARY MATERIAL}{SUPPLEMENTARY MATERIAL}
\section*{Supplementary Material}

Supplementary Methods are provided below.

\section*{Supplementary Methods}

\subsection*{Dataset Inclusion and Preprocessing Details}

For all datasets, we retained scans with matched phenotypic records, available extracted time series, complete age information, and passing post-extraction quality control. Subject identifiers were standardized with predefined regular expressions before matching phenotype and imaging records.

For COBRE~\cite{aine2017multimodal}, we loaded subjects from the available preprocessed release, extracted atlas time series, excluded the 295.70 bipolar type and 295.70 depressed type diagnosis subtypes, and then applied the common quality control filter. For ADNIDOD~\cite{weiner2014effects}, we identified available subject/session rs-fMRI files and matched them to phenotype records by SubjectID and Session. For Cam-CAN~\cite{shafto2014cambridge}, we included subjects with an available resting-state fMRI file and matched participant metadata. For OASIS-3~\cite{lamontagne2019oasis}, we matched subject/session rs-fMRI files to phenotype records after session harmonization. For ADNI~\cite{jack2008alzheimer}, we matched available subject/session rs-fMRI files to phenotype records by SubjectID and Session.

For ABIDE I~\cite{di2014autism}, we started from 1,112 rows in the ABIDE I preprocessed phenotypic table. Using the local ABIDE Preprocessed Connectomes Project cache with Configurable Pipeline for the Analysis of Connectomes (CPAC) \texttt{nofilt\_noglobal} outputs, accessed through \texttt{nilearn.fetch\_abide\_pcp}, 871 subjects had both a preprocessed rs-fMRI file and matching phenotypic metadata. After confound regression, Schaefer-100 time series extraction, and construction of the final benchmark table, 843 subjects were included. The 28 subjects present after PCP matching but absent from the final table are documented separately: 25 were Oregon Health \& Science University (OHSU) scans with only 78 raw volumes, and 3 were absent from the final processed table despite available PCP metadata and quality control fields.

Confound handling followed the files available in each release. For ADNIDOD, OASIS-3, and ADNI, we used the Nilearn simple confound strategy, including CompCor components from white matter and cerebrospinal fluid~\cite{behzadi2007component}; in the released implementation, these datasets also undergo global signal regression. For COBRE, Cam-CAN, and ABIDE, we used high variance confounds together with global signal regression~\cite{fox2009global}. Motion confounds were used as regressors during time series extraction but were not stored as benchmark metadata. For ADNIDOD, Cam-CAN, OASIS-3, and ADNI, the first five time points were removed before downstream analysis.

Final quality control used three criteria: final time series length of at least 100, covariance condition number between 10 and \(10^6\), and no regional time series column with exactly zero \(\ell_2\) norm. The lower cutoff of 10 was carried over from the original preprocessing pipeline. It is a heuristic for excluding OAS covariance estimates with little variation across eigenvalues, not a requirement for numerical stability. The processed benchmark tables include these time series length and covariance quality checks, but not scan-level motion measures or motion-based exclusion logs.

\subsection*{Connectome Construction}

We extracted cortical time series using the Schaefer-100 atlas~\cite{schaefer2018local}.

To obtain stable covariance estimates at the scan level, we used the Oracle Approximating Shrinkage (OAS) estimator~\cite{chen2010shrinkage}. OAS is a closed form shrinkage method that regularizes the sample covariance toward a scaled identity matrix. Covariance estimation was performed independently for each scan included in the analysis, so repeated scans from the same participant were treated as separate observations at this stage rather than pooled before estimation. Specifically, given \(\{x_i\}_{i=1}^{n}\subset\mathbb{R}^{p}\) under a zero mean Gaussian assumption, the OAS estimator is
\[
\widehat{\Sigma}_{\mathrm{OAS}} = (1-\alpha)S + \alpha \frac{\mathrm{Tr}(S)}{p} I,
\]
where \(S = \frac{1}{n}\sum_{i=1}^n x_i x_i^\top\) is the empirical covariance matrix, \(I\) is the \(p \times p\) identity matrix, and \(\alpha \in [0,1]\) is given by the closed form OAS shrinkage rule rather than tuned by cross-validation.

After covariance estimation, we added a small fixed diagonal term, \(\epsilon I\) with \(\epsilon = 10^{-5}\), to the covariance matrix. The regularized covariance matrix was converted to a correlation matrix by variance normalization, where each entry is divided by the product of the two corresponding regional standard deviations. During this division, any standard deviation below \(\sqrt{\epsilon}\) was replaced by \(\sqrt{\epsilon}\) to avoid unstable values when a region has nearly zero variance. We then added the same \(\epsilon I\) term to the correlation matrix as the final numerical conditioning guard. The covariance condition number quality control criterion is evaluated before the diagonal term is added.

\subsection*{Model Implementation Details}

Let \(C \in \mathcal{S}_{++}^{p}\) denote one regularized SPD correlation connectome for a single scan. For each outer evaluation split, the tangent-space reference \(C_{\ast}\in \mathcal{S}_{++}^{p}\) is estimated from the training data only and then held fixed when transforming the corresponding evaluation data. A congruence transformation first expresses \(C\) in the coordinate system of the reference point \(C_\ast\),
\[
\tilde{C} = C_{\ast}^{-1/2}\, C \, C_{\ast}^{-1/2}.
\]
The Riemannian logarithm map at \(C_\ast\) admits the closed form
\[
\log_{C_{\ast}}(C) = C_{\ast}^{1/2}\,\log(\tilde{C})\,C_{\ast}^{1/2} \in T_{C_\ast}\mathcal{S}_{++}^{p},
\]
where \(\log(\cdot)\) denotes the matrix logarithm. In practice, we use the equivalent whitened tangent coordinates \(T = \log(\tilde{C})\). If \(\tilde{C}=U\Lambda U^{\top}\) is the eigendecomposition with \(\Lambda\) diagonal, then
\[
T = \log(\tilde{C}) = U\log(\Lambda)U^{\top},
\]
with the logarithm applied elementwise to the eigenvalues. The symmetric matrix \(T\) is then represented by its upper triangular entries, including the diagonal. Diagonal coefficients keep unit weight, whereas off diagonal coefficients are scaled by \(\sqrt{2}\), following the standard tangent vector convention for symmetric matrices. Ridge regression is implemented with scikit-learn~\cite{pedregosa2011scikit}. In both the single-dataset and pooled Tangent-Space Ridge experiments, tangent vectors are standardized by feature before Ridge regression, with the scaling fitted on the training split and then applied to the corresponding validation or test data. The outer test fold is not used when fitting this scaler.

In the reported single-dataset and pooled age benchmarks, Ridge regularization strength \(\alpha\) is selected with \texttt{GridSearchCV} inside each outer training split. The implementation requires the current outer training split to contain at least two subject groups before running grouped inner cross-validation; this condition is satisfied for all reported age benchmark splits. The estimator is a \texttt{Pipeline} consisting of \texttt{StandardScaler} followed by \texttt{Ridge}. The alpha grid is \(\{10^{-3},10^{-2},10^{-1},1,10,100,1000\}\), and model selection uses negative mean absolute error. Inner cross-validation uses \texttt{GroupKFold} within the outer training set, with groups defined by subject identifier, so repeated scans from the same subject cannot be split between inner training and inner validation folds. In pooled analyses, subject identifiers are dataset-prefixed before grouping. The outer test set is not used for alpha selection. Within each inner fold, \texttt{StandardScaler} is fitted only on the inner training subset and then applied to the inner validation subset. Ridge regularization is selected after tangent features are built using the reference point from the full outer training set. The reference point is not recomputed separately inside each inner validation fold. Thus, alpha selection uses only the outer training data, but the tuning procedure is not fully nested with respect to the tangent space reference.

The SPDNet implementation~\cite{huang2017riemannian,aristimunha2026spd} is built from the following layers.

\begin{itemize}
\item \textbf{BiMap.} A bilinear mapping transforms an SPD matrix \(C \in \mathcal{S}_{++}^{p}\) into an SPD matrix with lower dimension via
\[
\bar{C} = W^\top C W,
\]
where \(W \in \mathbb{R}^{p \times q}\) is a learnable projection matrix, typically with full column rank and \(q \le p\), ensuring that \(\bar{C}\) remains SPD.

\item \textbf{ReEig.} An eigenvalue rectification layer enforces positive definiteness and improves numerical stability. Given the eigendecomposition \(C = U \Lambda U^\top\), the output is
\[
\bar{C} = U\,\bar{\Lambda}\, U^\top,
\]
where \(\bar{\Lambda} = \max(\Lambda, \epsilon I)\), \(\Lambda\) is diagonal, \(\epsilon>0\) is a small threshold, and the maximum is applied elementwise to the eigenvalues. In all SPDNet experiments, the ReEig layers use a fixed eigenvalue floor of \(\epsilon = 10^{-4}\).

\item \textbf{LogEig.} A logarithmic mapping layer applies the matrix logarithm to an SPD matrix. For \(C = U \Lambda U^\top\),
\[
\log(C) = U \log(\Lambda) U^\top,
\]
where \(\log(\Lambda)\) denotes the elementwise logarithm of the eigenvalues. This operation can be interpreted as a log domain representation at the identity matrix, since the Riemannian logarithm at \(I\) reduces to the ordinary matrix logarithm, \(\log_{I}(C)=\log(C)\), for \(C\in\mathcal{S}_{++}^{p}\).
\end{itemize}

After the final LogEig operation, SPDNet represents the symmetric log domain matrix by its unweighted upper triangular entries, including the diagonal, before passing the vector to the multilayer prediction head. This differs from Tangent-Space Ridge, whose tangent vectors use unit weights on the diagonal and \(\sqrt{2}\) weights off the diagonal.

For reproducibility, all main SPDNet experiments use the same fixed training hyperparameters. Optimization is performed with Adam, learning rate \(10^{-2}\), weight decay \(0\), and mean squared error loss. The batch size is 1024 for training and evaluation in the reported manuscript configuration. Training runs for at most 100 epochs, with gradient clipping at max norm 1.0. Model selection is based on validation MSE within each fold. Specifically, 10\% of the outer training subjects or subject groups are reserved through a random validation split that respects groups and uses a fixed seed, preserving the same grouping by subject used for the outer split. Early stopping uses patience 10. If the stopping criterion is met, the model with the lowest validation MSE is used for evaluation. If training reaches the maximum epoch budget, evaluation uses the model obtained at the final epoch. The fully connected prediction head is shared across SPDNet variants and consists of Linear \(\rightarrow\) ReLU \(\rightarrow\) LayerNorm \(\rightarrow\) Dropout \((0.5)\) \(\rightarrow\) Linear, with hidden dimension 100. Original and harmonized SPDNet variants use the same training hyperparameters and the same outer cross-validation folds. During SPDNet training, the implementation omits the final incomplete mini batch when its size is smaller than the requested batch size. This corresponds to \texttt{drop\_last=True} in the released data loaders.

Dummy uses \texttt{DummyRegressor} from \texttt{scikit-learn} and predicts the mean age while ignoring the features. CorrVec uses the off-diagonal upper triangular entries of each correlation matrix, excludes the constant diagonal, and multiplies retained entries by \(\sqrt{2}\). In both the single-dataset and pooled CorrVec Ridge experiments, the features are standardized before Ridge regression with a scaler fitted on the training split only and applied to the corresponding validation or test data. CorrVec Ridge uses the same \texttt{GridSearchCV} procedure and alpha grid as Tangent-Space Ridge, and the required subject-group condition is satisfied for all reported age benchmark splits. In the released pooled analysis script, CorrVec is an available baseline but is not part of the convenience default algorithm list. Reproducing the pooled tables therefore requires including \texttt{corr\_ridge} explicitly together with SPDNet, Tangent-Space Ridge, and Dummy.

The SPDNet configuration ablation is defined with input SPD dimension \(P=100\), corresponding to the Schaefer-100 atlas. The four ablation settings correspond to quarter dimension, half dimension, one full rank block, and two full rank blocks in SPDNet. The quarterdim variant uses one BiMap\((100 \rightarrow 25)\) + ReEig block, followed by LogEig and the shared fully connected head. The halfdim variant uses one BiMap\((100 \rightarrow 50)\) + ReEig block. The full rank one block architecture, labeled one in the figures, uses one BiMap\((100 \rightarrow 100)\) + ReEig block. The full rank two block architecture, labeled two, uses two repeated BiMap\((100 \rightarrow 100)\) + ReEig blocks before LogEig and the same fully connected head. The implementation also applies an additional numerical guard equivalent to ReEig before LogEig to preserve positive definiteness after learned SPD transformations. In the single dataset script, LogEig is evaluated in double precision, whereas the pooled implementation follows the default precision used by the pooled training code.

\subsection*{Harmonization and Split Details}

For each outer split, the harmonization reference point \(C_{\ast} \in \mathcal{S}_{++}^{p}\) is computed from the training data only. For a regularized SPD correlation connectome \(C \in \mathcal{S}_{++}^{p}\), the tangent-space embedding is
\[
Z = \mathrm{vec}\!\left(\log_{C_{\ast}}(C)\right),
\]
where \(\log_{C_{\ast}}(\cdot)\) is the Riemannian log map at \(C_{\ast}\) and \(\mathrm{vec}(\cdot)\) denotes vectorization. ComBat is fitted on the training tangent features, using dataset or site identity as the batch effect and age as the biological covariate. Diagnosis, sex, motion, and scan quality variables are not included. Empirical Bayes estimation is enabled, and age is modeled with a smooth term bounded to \texttt{(0,120)}, following harmonization work that preserves age-related variation in lifespan MRI data~\cite{pomponio2020harmonization}. The fitted harmonizer transforms associated evaluation data only through parameters learned from the training split, so no test observations are used when estimating \(C_{\ast}\), empirical Bayes parameters, or the age effect~\cite{marzi2024efficacy}. The interpretation caveat for using age as a covariate is given in the main text.

The corrected tangent features \(\bar{Z}\) are mapped back to the SPD manifold as
\[
\bar{C} = \exp_{C_{\ast}}\!\left(\mathrm{unvec}(\bar{Z})\right),
\]
where \(\exp_{C_{\ast}}(\cdot)\) is the Riemannian exponential map at \(C_{\ast}\) and \(\mathrm{unvec}(\cdot)\) reshapes the vector back to a symmetric matrix. For pooled GroupKFold, the test fold is transformed with the fitted training harmonizer and true test ages. For LODO, harmonization is fitted and applied only within the pooled training datasets; the held-out dataset is not transformed using its true ages and is not used to fit, refit, or adapt ComBat.

All experiments used fixed random seeds for the benchmark setup. A prespecified seed controlled stochastic steps during training, including train/validation splitting, fitting the regularized regression models, neural network initialization, and other stochastic model operations. An additional prespecified seed controlled shuffled sample order in the pooled benchmark before splitting. GroupKFold and LODO were deterministic once the benchmark tables were fixed. During SPDNet training, mini batches were randomly ordered for the training data, whereas validation and test data were evaluated in fixed order. This stochasticity was controlled at the run level rather than by assigning a separate random seed to the mini batch order in each fold. The source files also expose convenience defaults, including Cam-CAN as the default single dataset and batch size 100 in the single dataset script. These defaults are not the manuscript configuration, which uses explicit dataset and training arguments for the reported analyses.

\end{document}